# New studies of a variety of Southern Pulsating B stars

*Ulusoy, C., Engelbrecht, C.A. and Gülmez, T.*
*Department of Physics, University of Johannesburg, South Africa*
cerenu@uj.ac.za

**Introduction**

Recently, two independent instability regions have been defined among B-type stars in the HR Diagram. While one of them is the high-luminosity β Cephei instability strip, the other is the low- luminosity SPB instability strip. However, many β Cephei stars show hybrid features of various B-type pulsators as well (Balona et al. 2011). Consequently, pulsating B-type stars that have high luminosity and large radiation pressure may imply that we have to change our view of instability strips since the occurrence of this sort of hybridity is inevitable amongst these type of stars (Aerts, Christensen-Dalsgaard & Kurtz 2010). β Cephei stars are described as early B-type (B0-2.5) variables that are massive non-supergiants and show multiperiodicity of both low-order pressure and gravity modes with periods of several hours (Stankov & Handler 2005). They also have the same pulsation mechanism caused by the κ mechanism via Fe peak elements in partial ionozation driving zones that is common in B-type pulsators (Dziembowski & Pamyatnykh 1993).

We present preliminary results of multi-colour photometry of β Cephei stars observed in the Large Magellanic Cloud (LMC) - first studied by Pigulski & Kolaczkowski (2002) - and the Southern open cluster NGC 6200 identified in the All Sky Automated Survey-3 database (ASAS-3 - Pigulski 2005). Tentative identifications of pulsation modes have been made, and a number of new B pulsators have been noted. Interesting features have also been discovered in the light curves of some of these stars. These discoveries are expected to throw further light on the rotational behaviour and interior structure of B type stars.

**Observations of ALS 3721**

Observations for the star ALS 3721 (HD328862= NGC 6200 #4) located in the Southern open cluster NGC 6200 (see Figure1a) were carried out between the dates July 20th 2010 and May 15th 2011 by using the 75-cm and the 100-cm Cassegrain telescopes equipped with the UCTCCD and the STE3 CCD cameras in the Johnson U,B,V and I filters at the South African Astronomical Observatory (SAAO). The obtained data cover 15 days throughout both observing seasons. Some of the light curves of ALS 3721 show a remarkable standstill phase (see Figure 1b ) just before the its light maximum in the B,V and I filters.

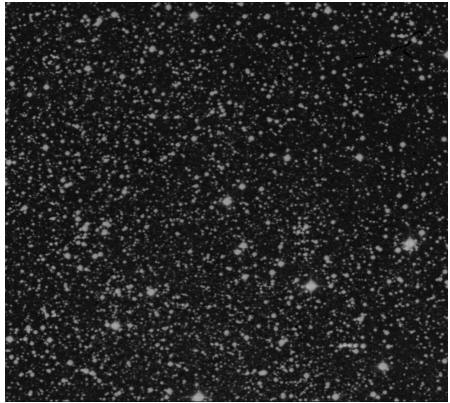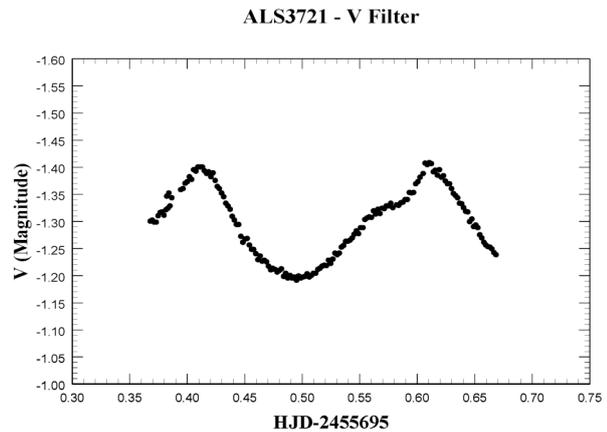

**Figure1a.** The location of ALS 3721 in the Galactic Open Cluster "NGC 6200"

**Figure1b :** Standstill phase of the light curve of ALS 3721 in the Johnson V Filter

## Observations of LMC Fields

Two fields surrounding stars previously identified as B pulsators in the LMC, *viz*. OGLE051841.98−691051.9 and OGLE052809.21−694432.1, were observed for a total of four weeks in December 2009 and January 2010 on the 100-cm Cassegrain telescope with the STE4 CCD camera at SAAO. The observing data covering 48.2 days were obtained in the Johnson U,B,V,R and I filters. Many other pulsators have been detected in the two LMC fields studied (one field is shown in Figure 2a). Analysis of the frequency spectra of these stars is in progress. Initial indications are that many low-frequency (in the 2–4 c/d regime) modes are present in the pulsators. In addition to these two LMC fields, two more LMC fields were studied in 2010-2011 and a further 10 fields in the LMC have been selected for future intensive photometric study.

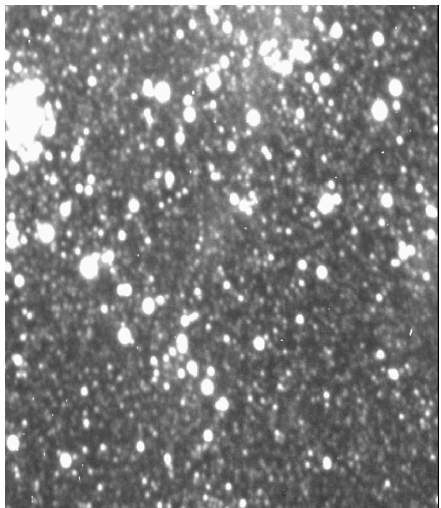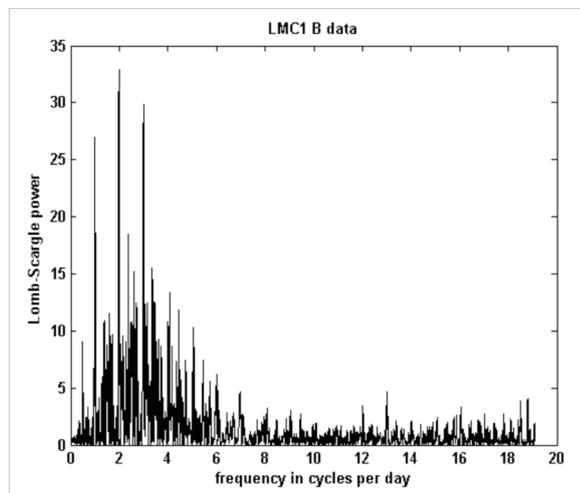

**Figure2a.** The location of OGLE 051841.98−691051.9(alias "LMC1") in the LMC

**Figure2b.** The Lomb-Scargle periodogram of B filter data for "LMC1". The strong signal at 2.01 c/d is an instrumental artefact

**Frequency Analysis**

Frequency analyses were performed by using Lomb-Scargle periodograms. Significance limits for the computed periodograms were determined by means of the Monte Carlo method discussed in Frescura et al. (2008). Only modes with a false alarm probability (FAP) below $10^{-5}$ (i.e. a statistical 4-sigma detection) for ALS3721 are discussed here. For LMC1, we discuss results with a false alarm probability (FAP) below $10^{-3}$ (a 3-sigma detection) due to the faintness of the target. The frequency spectra obtained in the Johnson B filter are presented in Figure 3 and 4 for ALS3721 and LMC1respectively.

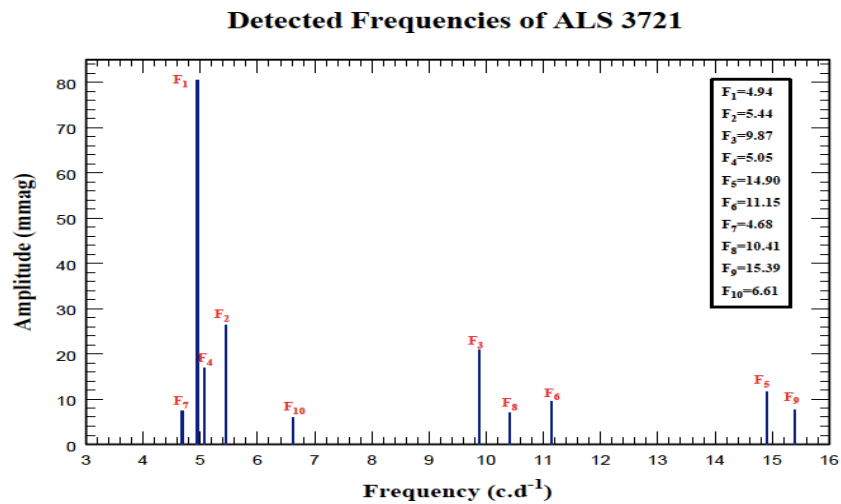

**Figure 3:** Detected frequency range of the star ALS 3721

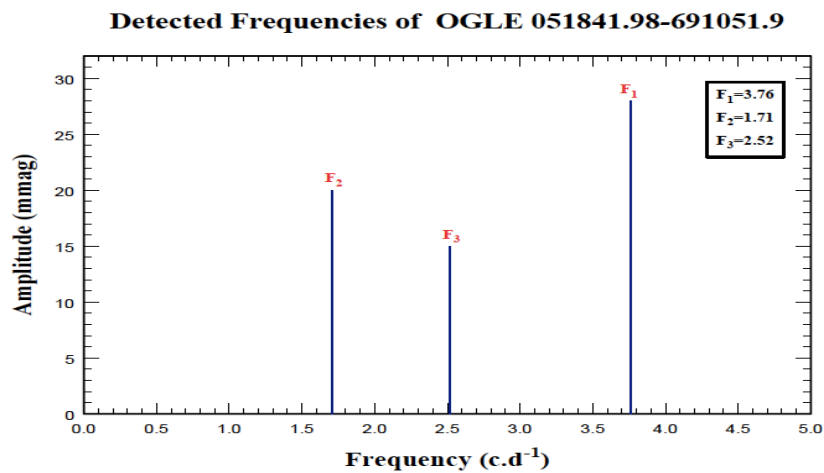

**Figure 4:** Detected frequency range of LMC1.

## Discussion

The star ALS 3721 was firstly classified as a Beta Cephei type star in the ASAS-3 survey published by Pigulski (2005). Additional to ASAS-3 survey analysis, we detected 8 more frequencies above the significance limit determined by our Monte Carlo method. We confirmed the main frequency and combination frequency given in the ASAS-3 survey. According to the frequency spectra of the star ALS 3721 showed in Figure 3, the detected periodicities fall in three groups on the frequency axis, with the largest group of 5 modes located in the classical Beta Cephei pulsation range but a triplet and a doublet located at surprisingly higher frequencies. Spectra of this star were also obtained via SALT. We are in the process of analysing these spectra and we strongly believe that they will yield more information about the structure of this particular star.

Miglio et al. (2007) presented B-star instability strips indicating that pulsation in low-metallicity (low-Z) stars will only be excited at the lower-frequency end of the spectrum. We indeed find three low-frequency modes in LMC1, which is highly likely to be a low-Z star. One of the modes (2.52 c/d) is probably a one-day alias of the 3.50 c/d mode reported by Pigulski & Kolaczkowski (2002), while the other 2 modes, at 1.71 c/d and 3.76 c/d agree with these authors' published results. Miglio et al. also indicated that higher-degree modes would be preferentially excited in low-Z stars. We look forward to testing this prediction through mode identification using our multi-colour photometry.

## Acknowledgements

We would like to thank the **South African National Research Foundation (NRF)** for supporting this project financially. We also thank the **South African Astronomical Observatory (SAAO)** for the allocation of observing time for these projects.